\documentclass[%
 reprint,
 amsmath,amssymb,
 aps,
prstab,
floatfix,
]{revtex4-1}

\usepackage{graphicx}
\usepackage{dcolumn}
\usepackage{bm}
\usepackage{textcomp, gensymb}
\usepackage[caption=false]{subfig}

\begin{document}

\preprint{APS/FEBE}

\title{Specification and design for Full Energy Beam Exploitation of the \\Compact Linear Accelerator for Research and Applications}

\author{E.W. Snedden}
    \email{edward.snedden@stfc.ac.uk}
\author{D. Angal-Kalinin}
\author{A.R. Bainbridge}
\author{A.D. Brynes}
\author{S.R. Buckley}
\author{D.J. Dunning}
\author{J.R. Henderson}
\author{J.K. Jones}
\author{K.J. Middleman}
\author{T.J. Overton}
\author{T.H. Pacey}
\author{A.E. Pollard}
\author{Y.M. Saveliev}
\author{B.J.A. Shepherd}
\author{P.H. Williams}
\affiliation{Accelerator Science and Technology Centre (ASTeC), STFC Daresbury Laboratory, Keckwick Lane, Warrington, WA4 4AD, United Kingdom}
\affiliation{Cockcroft Institute, Keckwick Lane, Warrington, WA4 4AD, United Kingdom}
 
\author{M.I. Colling}
\author{B.D. Fell}
\author{G. Marshall}
\affiliation{Technology Department, STFC Daresbury Laboratory, Keckwick Lane, Warrington, WA4 4AD, United Kingdom}

\date{\today}

\begin{abstract}
The Compact Linear Accelerator for Research and Applications (CLARA) is a 250~MeV ultra-bright electron beam test facility at STFC Daresbury Laboratory.  A user beam line has been designed to maximise exploitation of CLARA in a variety of fields, including novel acceleration and new modalities of radiotherapy.  In this paper we present the specification and design of this beam line for Full Energy Beam Exploitation (FEBE). We outline the key elements which provide users to access ultrashort, low emittance electron bunches in two large experiment chambers. The results of start-to-end simulations are reported which verify the expected beam parameters delivered to these chambers. Key technical systems are detailed, including those which facilitate combination of electron bunches with high power laser pulses. 
\end{abstract}

\maketitle

\section{\label{sec:intro}Introduction}
The Compact Linear Accelerator for Research and Applications (CLARA) is an ultra-bright electron beam test facility at STFC Daresbury Laboratory. The facility was conceived to test advanced Free-Electron Laser (FEL) schemes that could be implemented on existing and future short wavelength FEL facilities~\cite{Clarke2014}.

 The CLARA front-end, producing 50~MeV, 250~pC electron bunches from a 10~Hz S-band photoinjector gun and linac, was successfully commissioned in 2018~\cite{AngalKalinin2020}. Installation of accelerator modules to raise the beam energy to 250~MeV will be complete by the end of 2023. The front-end photoinjector gun will be replaced with a novel 100~Hz high repetition rate gun (HRRG)~\cite{McKenzie2014} which has been commissioned on an adjacent beam line. The remaining beam line consists of three 4 m S-band (2998.5~MHz) linacs, X-band fourth harmonic cavity (4HC) phase-space lineariser, dielectric dechirper, variable magnetic bunch compressor (VBC) and dedicated diagnostics line including a transverse deflecting cavity (TDC) for 6D phase space characterisation. The original CLARA concept included a laser heater (which can be installed in future, if required) and reserved space within the electron hall for a seeded FEL, including seeding laser, modulators, undulators and photon diagnostics; although the FEL has not been funded, the space has been reserved for future applications.

Beginning in 2018, access to electron beam from the CLARA front-end has been made available to users from academia and industry. This has enabled the testing of novel concepts and ideas in a wide range of disciplines, including the development of advanced accelerator technology~\cite{Pacey2019}, medical applications,~\cite{Small2021} and novel particle beam acceleration~\cite{Hibberd2020} and deflection~\cite{Saveliev2020,Pacey-PRAB2022} concepts.  

\begin{figure*}
\includegraphics[width=0.9\linewidth]{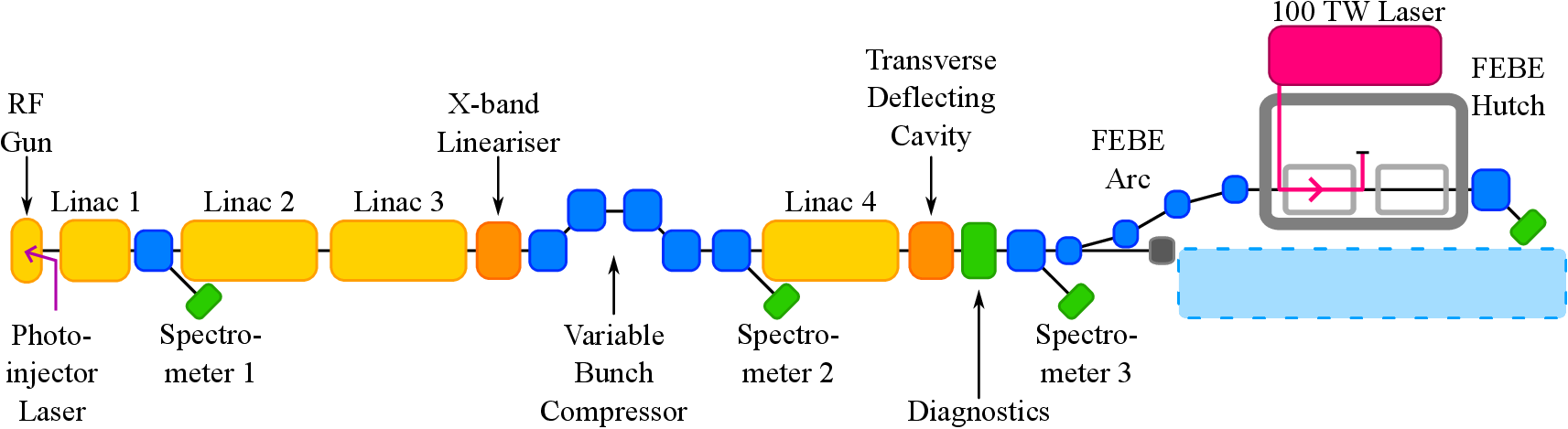}
\caption{\label{fig:CLARAbeamline}Schematic of the CLARA linear accelerator test facility, including the FEBE beam line, shielded FEBE hutch, 100~TW laser system, and space reserved for potential future applications (shaded blue area).}
\end{figure*}

Based on increasing user demand for access, a decision has been made to design and build a dedicated beam line for user applications at the full CLARA beam energy of 250~MeV.  As shown in Fig.~\ref{fig:CLARAbeamline}, the beam line for Full Energy Beam Exploitation (FEBE) will be installed parallel to the space originally allocated for an FEL.

There are only handful of test facilities worldwide which provide user access to mid-energy range (less than 300~MeV), high brightness electron beams to test proof of principle novel applications. A survey of beam dynamics challenges of such mid-energy high brightness facilities in Europe has recently been carried out and was presented at IPAC'23~\cite{Angal-Kalinin_IPAC2023}. In addition to CLARA, there are three other facilities in Europe; CLEAR@CERN~\cite{clear, clear1}, ARES@DESY~\cite{ares, ares12022}, and SPARC\_LAB@INFN~\cite{sparc_lab} in this energy range. 

The CLEAR facility provides bunch trains of beam energy up to 230~MeV at a maximum repetition rate of 10~Hz; the number of micro-bunches in each train can be varied from 1-150 with spacing of 1.5 or 3~GHz, and bunch charge can be varied from 5~pC to 3~nC. The ARES facility provides single bunches up to 160~MeV at a maximum repetition rate of 50~Hz, and bunch charge can be varied from 3~fC to 280~pC. The SPARC\_LAB facility provides single bunches up to 180~MeV at a maximum repetition rate of 10~Hz, and the bunch charge can be varied from 10~pC to 2~nC. In addition to CLARA, SPARC\_LAB is the only facility with access to a high power laser (FLAME~\cite{Bisesto2018-2}, 200~TW) to allow combined electron-laser experiments. 

A survey of CLARA stakeholders was performed to inform the design of the new beam line; the results of this survey identified three key design principles: 
\begin{enumerate}
    \item FEBE should provide access to a dedicated shielded experiment area (`hutch'), accessible to users without switching off CLARA.
    \item The hutch should incorporate large experiment chambers compatible with a wide range of possible experiments.
    \item The beam line should allow the synchronized interaction of electron bunches with a high-power ($\sim$100~TW) laser.
\end{enumerate}

The decision to provide a dedicated shielded hutch was taken following consultation with other medium-energy accelerator facilities. This arrangement allows on demand user access to the experimental area without fully switching off the accelerator, which reduces disruption, improves machine stability, and allows experiments to resume promptly after access. This type of access is not currently possible at other similar facilities in Europe, although CLEAR has developed robotic systems to minimise user access requirements during some types of experiment.

In this article we report on the specification and design of the FEBE beam line, which is currently under construction and will begin commissioning in 2024. The article is broken down as follows: the layout of the machine and beam specification is presented in Sec. \ref{sec:layout}; Sec. \ref{sec:s2e} reports the results of beam dynamics simulations from the CLARA photoinjector through to the FEBE beam dump; and Sec. \ref{sec:tech} details the key technical accelerator systems expected to underpin future user exploitation. The article concludes with a summary in Sec. \ref{sec:summary}.

\section{\label{sec:layout} Layout and beam specification}
\begin{figure*}
\includegraphics[width=1\linewidth]{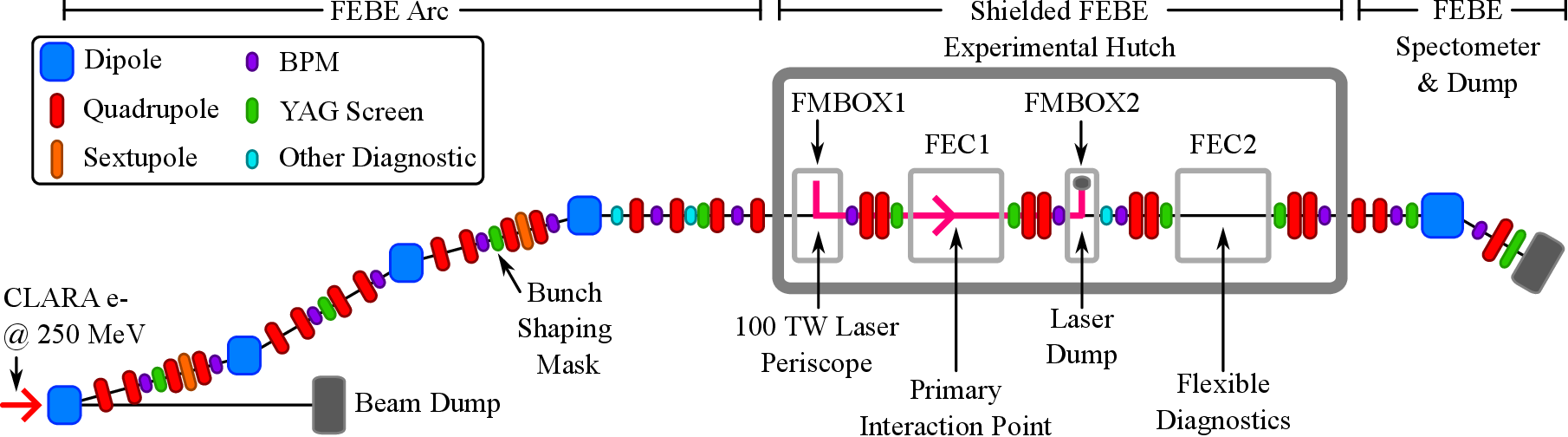}
\caption{\label{fig:FEBElayout}Schematic of the new beam line for Full Energy Beam Exploitation (FEBE) by users, including arc (connecting to the upstream CLARA main beam line, Fig.~\ref{fig:CLARAbeamline}), experiment hutch, and post-hutch beam dump with energy and emittance diagnostics.}
\end{figure*}

FEBE has been designed to support a variety of experiments across the fields of accelerator applications and accelerator technology. A user survey performed in 2018 established a particular interest in the exploitation for novel acceleration R\&D including: external injection of electron bunches into a plasma accelerator stage (using both beam and laser-driven configurations); structure wakefield acceleration, encompassing use of metallic, dielectric and novel (e.g. metamaterial or photonic crystal based) structures; and dielectric laser acceleration, including both direct optical laser coupling to a solid structure or prior conversion to longer wavelength (THz-band). A schematic of the beam line is shown in Fig.~\ref{fig:FEBElayout}; the position of the beam line within the CLARA facility is shown in Fig.~\ref{fig:CLARAbeamline}.

The requirements of novel acceleration techniques have been identified as the most challenging of the anticipated user requests, and has been used to drive the FEBE beam specification and underpinning accelerator technology (the latter outlined in Sec. \ref{sec:tech}). Characteristics of electron drive beams required for novel acceleration include: high charge and peak currents (order 1~kA)~\cite{Habib2023}, short bunch lengths (order 10~fs), and small transverse beam sizes ($<$10~$\mu$m)~\cite{Wu2021}. More demanding experiments and applications may require a combination of multiple characteristics simultaneously~\cite{Emma2021}. To verify and optimise the various acceleration techniques, diagnostics for the characterisation of electron bunches both before and after interactions are required~\cite{Downer2018}.

FEBE also expects to host a variety of target irradiation experiments, including R\&D in Very High Electron Energy (VHEE) therapy, radiation generation, and electron detectors. This is a broad category and experiment chambers must be suitably flexible to accommodate a wide variety of samples (both vacuum and in-air), with suitable motion control for accurate positioning. Diagnostics to validate beam parameters delivered to target are also required.

A schematic of the FEBE beam line is shown in Fig.~\ref{fig:FEBElayout} and is broken down into three sections: an arc and matching section connected to the main CLARA beam line; the FEBE experiment hutch, which brings the electron beam to a focus at two possible interaction points (IPs); and post-hutch transport line and beam dump. The FEBE experiment hutch is a $10\times5.4\times3$~m\textsuperscript{3} dedicated area for users to perform electron beam experiments.

The hutch is transversely offset from the main CLARA beam line using a FODO structure to provide -I transform between two dipoles (dipole angles of $14\degree$), and optimised to minimise emittance growth due to Coherent Synchrotron Radiation (CSR).\cite{DiMitri2013} This solution leads to a strong focusing, achromatic and non-isochronous arc with large natural second-order longitudinal dispersion, requiring correction by sextupole magnets at positions of high dispersion. Six quadrupole families allow matching to the main beam line for a range of electron beam configurations. The arc has a static $R_{56}$ value of 7.7~mm with no residual dispersion. Longitudinal bunch compression of the electron beam within the hutch can be achieved using a combination of the FEBE arc and the upstream VBC.

\begin{figure}[b]
\includegraphics[height=3.8cm]{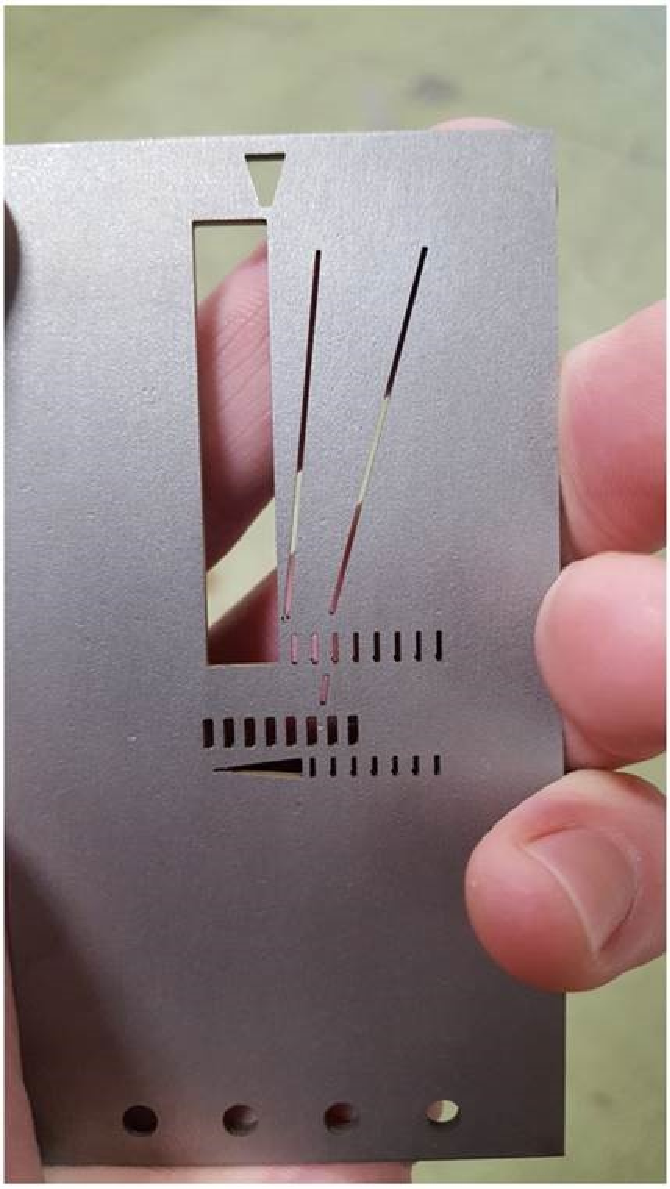}
\includegraphics[height=3.8cm]{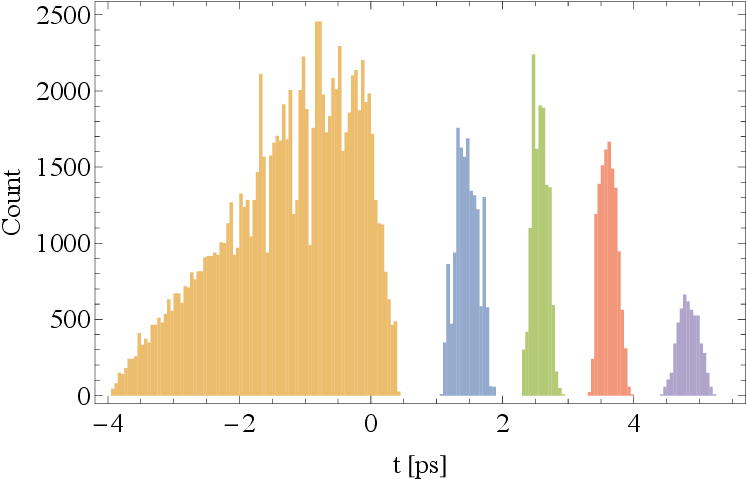}
\caption{\label{fig:mask}Left: photograph of the mask to be installed in FEBE for variable longitudinal shaping. Right: simulated 250~pC longitudinal profile at the FEBE FEC IP using the bottom hole of the mask, including triangular drive bunch ($\sim$1~ps RMS) and four trailing bunches ($\sim$100~fs RMS duration with 1~ps separation).}
\end{figure}

The FEBE arc includes a mask array positioned at a point of high dispersion for shaping of the bunch longitudinal distribution, including generation of: drive/main bunch-pairs with variable delay; a single ultra-short (order 1~fs duration) low charge bunch; and, as shown in Fig.~\ref{fig:mask}, a drive bunch with a train of witness bunches. The mask is made from 5~mm tungsten and can be changed to meet user requirements. Alternatively, multiple electron bunches can be generated at the photocathode via manipulation of the photoinjector laser, before acceleration and transport to FEBE.

The beam transport is designed to deliver a strong focus to two possible IPs (IP1/2), each located within a large-volume ($\sim$ 2~m\textsuperscript{3}) experiment chamber designated FEBE Experiment Chamber (FEC) 1/2. The double-IP design provides flexibility in experiment design and implementation. For example: the interaction between the electron beam and laser generated in FEC1 can be captured and probed with beam diagnostics installed in FEC2. The design also allows multiple independent experiments to be installed in FEC1 and FEC2 where compatible, minimising downtime for setup. 

\begin{table*}[t]
\caption{\label{tab:tablebp}
FEBE beam parameters at the FEC1 IP. All beam parameters are specified for 250~MeV. Symbol $\sigma_{i}$ indicates RMS value; $\epsilon_{N,i}$: normalised projected RMS emittance. Parameters for initial commissioning and those targeted thereafter following required periods of machine development are presented. 
}
\begin{ruledtabular}
\begin{tabular}{lcccc}
 & \multicolumn{2}{c}{Commissioning} & \multicolumn{2}{c}{Machine Development} \\
Parameter & High Charge & Low Charge & High Charge & Low Charge \\
\hline
Charge (pC) & 250 & 5 & 250 & 5 \\
$\sigma_t$ (fs) & 100 & 50 & $\leq 50$ & $\ll 50$ \\
$\sigma_x$ ($\mu$m) & 100 & 20 & 50 & $\sim 1$ \\
$\sigma_y$ ($\mu$m) & 100 & 20 & 50 & $\sim 1$ \\
$\sigma_E$ (\%) & $<5$ & $<1$ & 1 & 0.1 \\
$\epsilon_{N,x}$ {[}$\mu$m-rad{]} & 5 & 2 & $<5$ & $<1$ \\
$\epsilon_{N,y}$ {[}$\mu$m-rad{]} & 5 & 2 & $<1$ & $<1$
\end{tabular}
\end{ruledtabular}
\end{table*}

To meet novel acceleration requirements, FEC1 includes the capability to combine electron beams with high power lasers at the IP. The laser is introduced into the beam line in a dedicated mirror box, whereupon it can be transported directly to FEC1, or focused and co-propagated (using a $\sim$3.5~m effective focal length off-axis parabola) with the electron beam to the IP. A second transport line bypassing the off-axis parabola and connecting to FEC1 allows shorter focusing geometries to be generated directly within the chamber. A downstream mirror box is used to separate and dump the laser following interaction. Further detail on the laser system and laser transport is given in Sec. \ref{sec:tech}. Apertures throughout the beam line must accept both the high power laser and the electron beam. This includes, for example, the four quadrupoles around FEC1, which must be both large aperture (radius $\sim$68~mm) and high gradient to achieve a tight focus at the IP.

Some experiments (e.g., novel acceleration including plasma and dielectric) will aim to produce beams of higher energy than 250~MeV. All post-FEC1 magnets are therefore specified for up to 600~MeV, allowing high energy beam capture measurements in FEC2 and transport to the beam dump.

The post-hutch beam line is designed to provide transport to the beam dump housed in the main CLARA accelerator hall. A 20\degree dipole magnet is used to bend the beam to a large aperture Yttrium Aluminium Garnet (YAG) scintillation screen for beam imaging and energy spectrometry.  The dispersion at the spectrometer YAG station is modified by a single quadrupole specified to achieve the zero-dispersion condition at the screen position. To optimise the beam imaging, the horizontal and vertical beta-functions are minimised at the YAG location, to maximise the energy resolution whilst also maximising the image intensity; this is achieved by utilising post-IP hutch quadrupoles as part of the dump line matching.

Access to the hutch with the accelerator running is made possible via interlock of the FEBE arc dipoles to the machine personal safety system.  The total beam power within the hutch is limited to 6.25~W, which offers sufficient flexibility with available bunch charge (maximum 250~pC), bunch repetition rate (maximum 100~Hz), $\sim$100~TW laser repetition rate (5~Hz) and final beam energy (250-2000~MeV).  Radiation shutters on either side of the enclosure (in the CLARA tunnel) are used to shield the hutch from radiation generated from the main CLARA accelerator.

The beam specification at the FEC1 IP is presented in Tab. \ref{tab:tablebp} for four possible accelerator configurations. The Commissioning target high and low charge modes will form the baseline parameters made available to users for the initial beam exploitation period. Progression towards more demanding parameters including higher peak current and improved transverse quality will be made through periods of machine development. Machine development will also include development of appropriate diagnostic systems required to verify those parameters.\cite{Pacey:2022rea}

\section{\label{sec:s2e} Start-to-end simulations}
Start-to-end simulations were performed to evaluate and optimise the electron beam properties at the FEC1 IP. Simulations targeted machine configurations that deliver one or more beam properties relevant to the anticipated user experiments, including high peak current and charge density.  Table \ref{tab:tablebp} details the four main accelerator operating modes addressed through simulations.

Particle tracking simulations were carried out using ASTRA~\cite{Floetmann}, ELEGANT~\cite{Borland2000} and GPT~\cite{GPT}, accounting for the non-linear effects (both longitudinal and transverse) of space charge and CSR. The simulation codes were accessed via a python-based framework (SimFrame) developed at STFC Daresbury Laboratory, which allows a single human-readable lattice file to be deployed consistently across several codes. 

ASTRA and GPT were primarily used to simulate the CLARA front end at low energy (below 35~MeV), where transverse and longitudinal space charge forces are the dominant emittance-diluting processes. Tracking through the injector with 256,000 macroparticles was found to produce good agreement between codes within a reasonable computation time. 

Above 35~MeV, ELEGANT was primarily used due to its processing speed and the inclusion of CSR effects in the bunch compressor and FEBE arc. These high-energy sections (above 35~MeV) of the machine were simulated with 32,000 macroparticles (down sampling the output of the injector simulations); high-fidelity simulations used for machine optimisation required 256,000 macro-particles. Extensive comparisons were made between ELEGANT and ASTRA at higher energies, showing small differences due to transverse space-charge effects (not included in ELEGANT) but similar longitudinal space-charge forces. ASTRA does not include adaptive space-charge meshing, which complicates particle tracking under bunch-compression scenarios.

Bunch properties (for both low and high charge configurations in Tab.~\ref{tab:tablebp}) at the FEC1 IP were inferred from the statistical distribution of particles from tracking simulations. To characterise the longitudinal profile of the bunch, the peak current, current full-width at half (FWHM) and quarter max (FWQM), and charge fraction (integrated charge within the current FWQM) were extracted. Due to the statistical nature of particle tracking, particularly with space charge and CSR effects, a smoothing algorithm (Kernel Density Estimator) was applied to the longitudinal charge distribution. The probability density function and associated cumulative density function were obtained and used to evaluate the full-width and charge fraction values. Each configuration was tuned to maximise the peak beam current, which was sensitive to the length of each slice in the time domain and average number of macro-particles per slice. 

\begin{figure}[t]
\includegraphics[width=8.6cm]{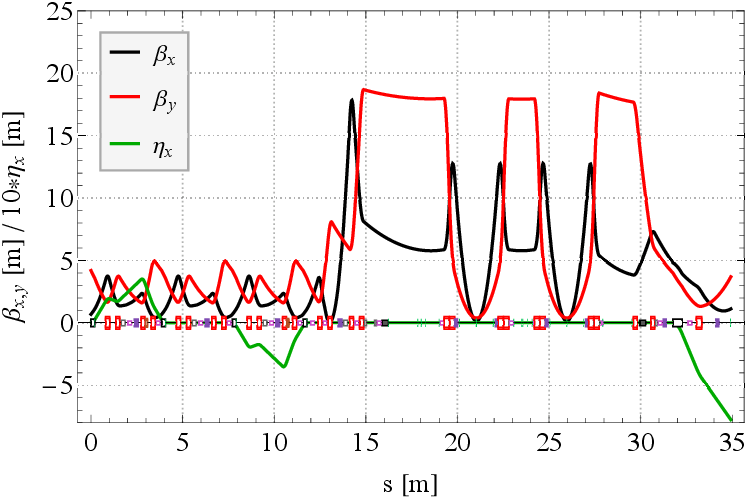}
\caption{\label{fig:twiss} Twiss parameters for the full FEBE beam line starting from the first dipole in the FEBE arc.}
\end{figure}

For high peak current generation, the most effective accelerator actuators were those modifying the longitudinal phase space of the bunch: the amplitudes and phases of radio-frequency (rf) linacs, wakefield structures (including dielectric energy dechirping), and transverse bending structures which lead to coherent effects like CSR.  The FEBE beam line contains only one of these structures, in the form of the FEBE arc and its associated CSR effects, where the arc is designed to minimise the induced CSR kicks, as described in~\cite{DiMitri2013}. In principle, the $R_{56}$ of the arc can be adjusted by varying the strengths of the two quad families allowing $R_{56}$ values between $\pm$~20mm; this however produces non-zero dispersion at the exit of the arc. The Twiss parameters for the nominal FEBE beam line are shown in Fig.~\ref{fig:twiss}.

\begin{table}[b]
\caption{\label{tab:cons}
Optimisation constraints used for particle tracking.
}
\begin{ruledtabular}
\begin{tabular}{cc}
\textrm{Constraint}&
\textrm{Value}\\
\colrule
Linac gradients & $<$25 MV/m\\
Beam energy & 240-260 MeV\\
Peak current @ IP & $>$2.5 kA\\
Slice $\epsilon_{N,x}$ @ IP & $<$10 $\mu$m-rad\\
Slice $\epsilon_{N,y}$ @ IP & $<$1 $\mu$m-rad\\
FWQM & 0.03 ps\\
FWQM charge fraction & $>$75\%
\end{tabular}
\end{ruledtabular}
\end{table}

The main optimisation actuators for the FEBE beam are found in the preceding CLARA beam line, shown in Fig.~\ref{fig:CLARAbeamline}. The first two-metre S-band injector linac (Linac~1) can act as either a standard low energy accelerating structure or a longitudinal bunching structure for short single-spike operation. The remaining three four-metre long S-band linacs (Linacs~2-4) provide acceleration up to a nominal beam energy of 250~MeV. A chicane-type VBC is located between Linac~3 and Linac~4, with a X-band 4HC immediately before the VBC for longitudinal phase space curvature compensation. The VBC is located at a nominal energy of $\sim$180~MeV, to maximise its effectiveness for the moderate compression required for the original CLARA FEL concept.

The nominal FEBE arc $R_{56}$ of $+8$~mm is of the opposite sign from the main bunch compressor $R_{56}$ of $-42$~mm. For standard operating modes the FEBE arc is decompressing for a standard longitudinal chirp. Reversing the chirp using Linac~4 after the VBC is difficult; achieving maximum compression in the FEBE hutch therefore requires over-compression in the VBC, followed by re-compression during transport in the FEBE arc. This has the additional benefit of reducing the CSR effects in the FEBE arc itself, at the cost of a longitudinal cross-over in the VBC.

The CLARA beam line also contains a dielectric-based dechirper structure used to minimise the projected energy spread at the FEL~\cite{Colling2019}. As shown in Fig.~\ref{fig:FEBElayout}, this device is located a few metres upstream of the FEBE extraction dipole. The dechirper was modelled in ELEGANT as a 1D longitudinal wakefield element using a theoretically calculated Green's function, which has been experimentally verified for this structure~\cite{Pacey-PRAB2022}. Future work will incorporate full 3D simulations of the wakefield dynamics. 

Three different algorithms were used for optimisation of the combined CLARA FEBE beam line: a SciPy-based genetic algorithm, a SciPy-based Nelder-Mead simplex algorithm, and a custom-written Nelder-Mead simplex algorithm.  The optimisation constraints are shown in Tab. \ref{tab:cons}.  At high peak currents CSR effects can drastically increase the horizontal projected emittance, making beam transport through narrow experimental apertures difficult and restricting the minimum beam focus that can be achieved at each IP. Emittance growth was therefore introduced as an additional constraint.

\begin{figure}[t]
    \subfloat[\label{fig:pdf:a}]{%
    \includegraphics[width=\linewidth]{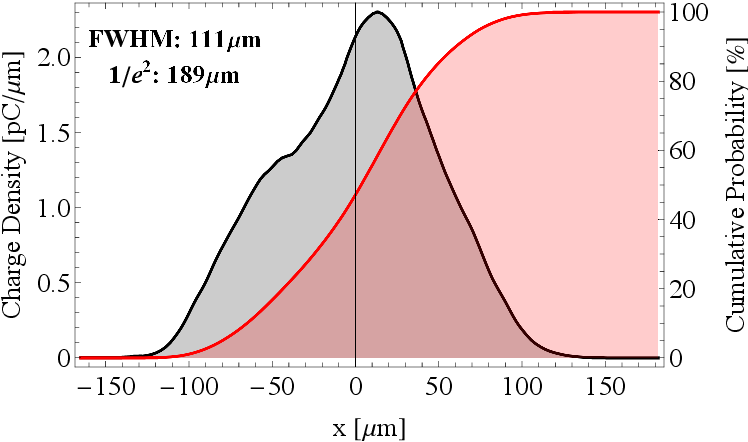}%
    }\hfill
    \subfloat[\label{fig:pdf:b}]{%
    \includegraphics[width=\linewidth]{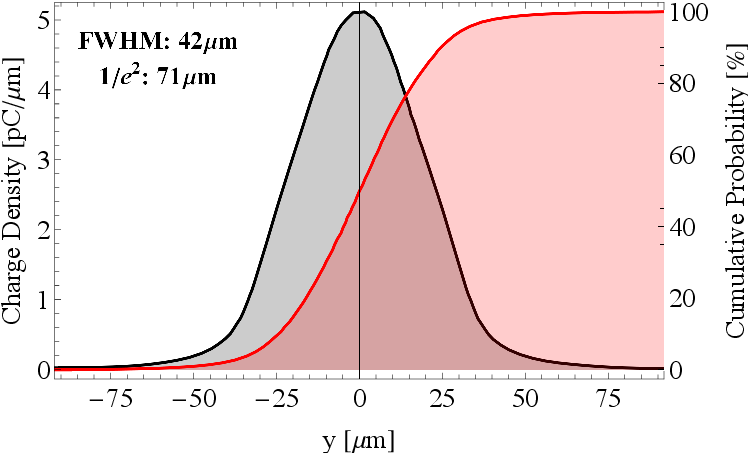}%
    }\hfill
    \subfloat[\label{fig:pdf:c}]{%
    \includegraphics[width=\linewidth]{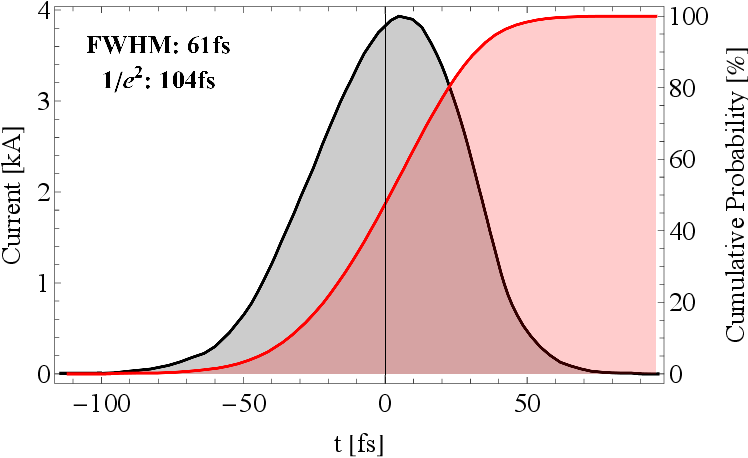}%
    }\hfill
    \caption{\label{fig:pdf}Charge densities and cumulative probability functions for an optimised 250~pC solution given at the FEBE FEC1 IP, for (a) the transverse horizontal, (b) vertical and (c) longitudinal planes. Beam parameters extracted from the density functions are presented.}
\end{figure}

Optimisation of the transverse lattice parameters was found to vary with rf focusing from Linacs~1-4. Lattice matching was not performed on an iteration-by-iteration basis, but re-matching conducted when the tracked beta functions strayed too far from design values. In general, re-matching was only performed between optimisation sessions. Figure \ref{fig:pdf} shows projections of the bunch charge distribution at the FEC1 IP, calculated for the nominal simulation of the high charge (250~pC) operating mode. We utilize low bandwidth KDE functions to smooth the noisy longitudinal charge distribution and present both the linear charge density and cumulative density functions. Relevant length measurements are shown for each plane, indicating the spatial and temporal dimensions of the distributions.

To evaluate the possibility of micro-bunching of the bunch longitudinal phase space, a semi-analytic model~\cite{Brynes2017} was developed to compute the micro-bunching gain and energy modulation.  This model computes the longitudinal space charge and CSR impedance in the drift spaces, linacs and bunch compressors, using simulated beam parameters at various points. It allows for intra-beam scattering (IBS) effects to be included or excluded to demonstrate the potential impact on the damping of micro-bunching gain.

The micro-bunching gain was calculated in stages with CLARA separated into sections. The longitudinal space charge-induced energy modulation was calculated for each linac and long drift section iteratively, as this parameter depends on the beam energy and beam size.  Average values for the transverse beam size were used for each machine section and a linear increase in beam energy applied for the linacs. The final bunching factor at the exit of the VBC (as a function of uncompressed modulation wavelength) was then given by the impedance in the integral summed over all preceding sections. This process was repeated for the remaining linacs, drifts and the arc compressor.

Micro-bunching gain is highly dependent on the uncorrelated energy spread ($\sigma_{E,0}$), which causes the exponential damping of modulations.  ASTRA simulations of the HRRG and Linac~1 were used to determine $\sigma_{E,0}=1.06$~keV.  The gradient and phase of the gun and Linac~1 were set to ($-9\degree$, 120~MV/m) and ($-16\degree$, 21~MV/m) respectively. The value of $\sigma_{E,0}$ was calculated as the mean value between $\pm2$~ps.  A full summary of the main lattice parameters used for the calculation of the micro-bunching gain factor is shown in Tab. \ref{tab:mb}.

\begin{figure}[t]
\subfloat[\label{fig:ibs:a}]{%
    \includegraphics[width=0.9\linewidth]{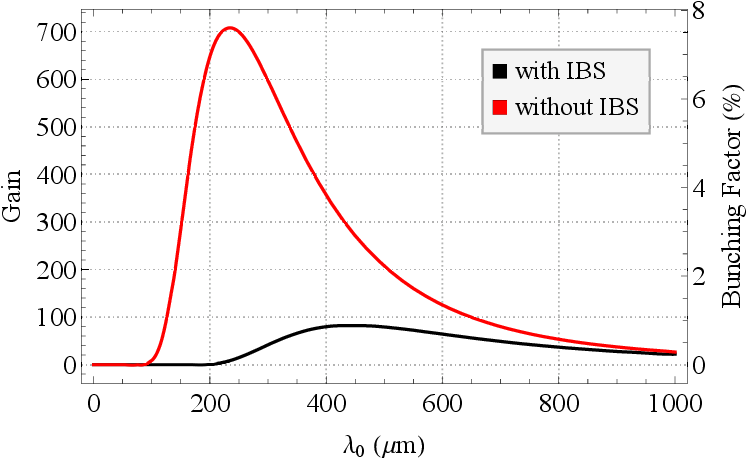}%
    }\hfill
\subfloat[\label{fig:ibs:b}]{%
    \includegraphics[width=0.825\linewidth]{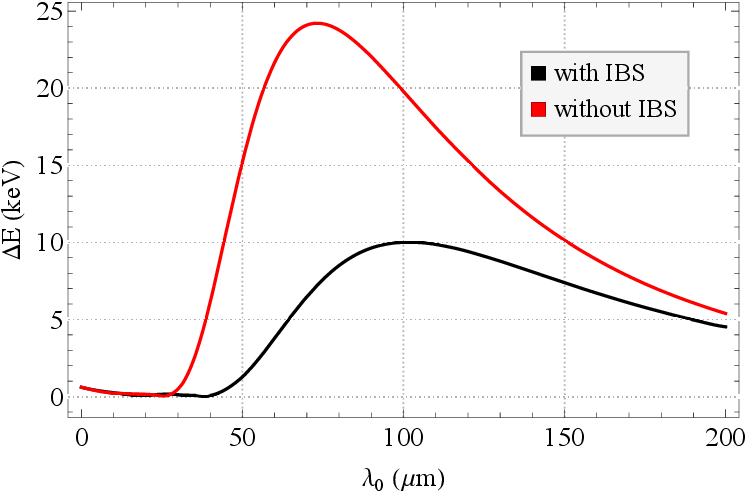}%
    }\hfill 
\caption{\label{fig:ibs} Semi-analytic model predictions of the (a) micro-bunching gain and (b) energy modulation amplitude as a function of initial modulation wavelength ($\lambda_{0}$) after compression in both the VBC and the FEBE arc, with and without the energy spread added by intra-beam scattering (IBS).}
\end{figure}

The output from the semi-analytic model is shown in Fig.~\ref{fig:ibs:a}, which shows the micro-bunching gain as a function of the initial modulation wavelength and the final bunching factor, after compression in both the variable bunch compressor and the FEBE arc.  Even without including the damping effect caused by IBS, the maximum final bunching factor is around 7\%.  This level of bunching in the longitudinal plane is expected to be tolerable for most FEBE experiments.  The final bunching factor drops by around a factor of ten when IBS is included.

The energy modulation amplitude is shown in Fig.~\ref{fig:ibs:b}. The damping caused by scattering is only around a factor of two at the level of maximum energy modulation, but the calculation without scattering gives this maximum value at approximately 0.1\% of the average beam energy (240~MeV).

\begin{table}[b]
\caption{\label{tab:mb}
Beam and lattice parameters used for the computation of the micro-bunching gain. Beam parameters as measured after Linac~1, taken from simulation.
}
\begin{ruledtabular}
\begin{tabular}{ccc}
\textrm{Parameter}&
\textrm{Value}\\
\colrule
Bunch charge [pC] & 250\\
Initial beam energy [MeV] & 35\\
Final beam energy [MeV] & 240\\
Initial uncorrelated energy spread [keV] & 1.06\\
Initial bunch length [ps] & 2.42\\
Initial peak current [kA] & 0.05\\
Normalised emittance [$\mu$m-rad] & 0.42\\
VBC compression factor & 10\\
Arc compression factor & 4\\
\end{tabular}
\end{ruledtabular}
\end{table}

\section{\label{sec:tech} Accelerator technology}
\subsection{Beam diagnostics}

Beam diagnostics for FEBE are designed to verify to the targeted commissioning beam parameters as defined in Tab.~\ref{tab:tablebp} and the requirements of anticipated experiments. The  beam line includes an array of well developed or commercially available diagnostic systems including: Ce:YAG screens, strip-line beam position monitors, integrated current transformers, and Faraday cup.

The most demanding requirements on diagnostic systems are set by the beam parameters generated in novel acceleration techniques, including measurement of micrometer-scale transverse profiles, 10~fs bunch duration, emittance and broadband energy spectra at high resolution. Shot-by-shot characterisation is required, particularly in techniques were inherent instabilities may manifest variation of the output beam parameters; non-invasive diagnostics are required for control and optimisation. Meeting these challenges has required a dedicated diagnostics R\&D programme, key elements of which are detailed below.

\subsubsection{High dynamic range charge measurement}
The flexible beam delivery of FEBE will require both accurate and precise measurement of charges from $<$5~pC to 250~pC. Low charge machine setups will be important for the initial phases of some novel acceleration experiments, as the transverse spot sizes and bunch lengths will be significantly lower than at high charge, as shown in Tab.~\ref{tab:tablebp}. A new electronics front end has been developed for the Faraday cups on CLARA which will maintain accuracy and a high signal-to-noise ratio across this charge range. This system is comprised of three components: an analog signal chain which converts signals from charge devices into an output pulse proportional to charge; a charge injection circuit for onboard calibration of the analog signal chain; and digital control circuitry to enable operators to adjust the settings of the analog front end. Further details are given in \cite{Mathisen2022}.

\subsubsection{Spectrometer dipole}
An in-vacuum permanent magnet dipole spectrometer has been designed to measure the energy spectra of beams generated in novel acceleration experiments. The spectrometer can measure beam energies in the range 50-2000~MeV to accommodate both beams with significant shot-to-shot instabilities and bunches with high energy spread. The spectrometer will nominally be installed in FEC2 as per experiment requirements.

\begin{figure}[b]
\includegraphics[width=0.9\linewidth]{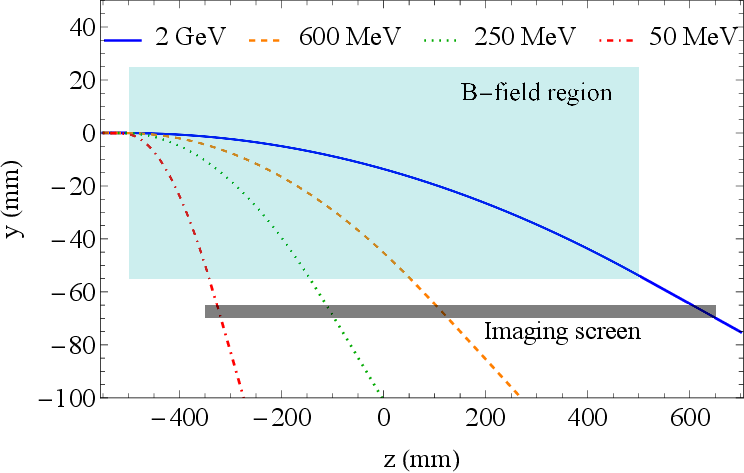}
\caption{\label{fig:trajectory}Plot of the predicted trajectories of electrons in the full energy range through the 1~m long five-section spectrometer. $z=0$ is located at the spectrometer mid-point, $y=0$ is the beam entry height. The region of magnetic flux is shown in blue between $\pm500$ mm; trajectories for 2~GeV (solid blue), 600~MeV (orange dashed), 250~MeV (green dotted) and 50~MeV (red dash-dot) are plotted.}
\end{figure}

\begin{figure*}[t]
\includegraphics[width=\linewidth]{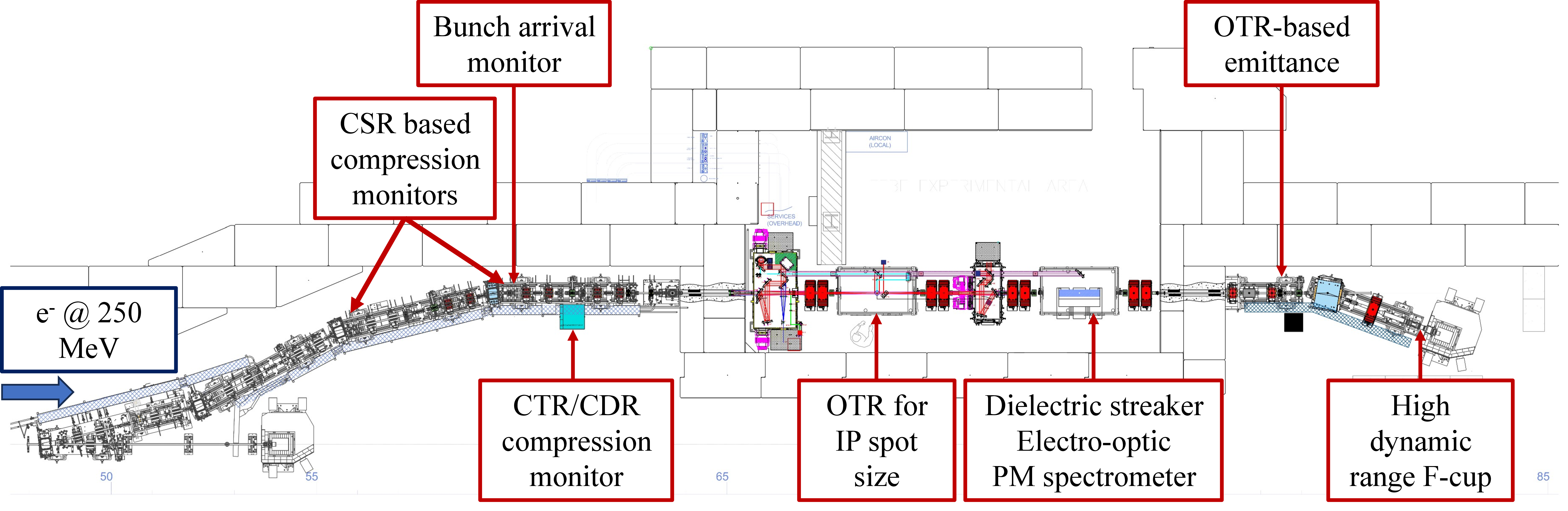}
\caption{\label{fig:diagnostics} Engineering diagram of FEBE beam line, highlighting areas with diagnostic systems undergoing R\&D. These
areas are supported by standard beam diagnostics including Ce:YAG screens, beam position monitors and integrated current transformers.
}
\end{figure*}

A long C-core dipole disperses electrons of different energies via the open side onto a long screen. The screen is positioned just below the spectrometer and angled at 45\degree to allow viewing by cameras positioned in air. The design is modular, consisting of up to five identical sections with lengths of 200~mm which co-locate using precision metal dowels. All five sections (total magnet length of 1000~mm) will be used for GeV-scale experiments. A shortened version consisting of three segments at 600~mm total length may be employed for experiments at lower energy gain ($\leq 600$~MeV). 

The magnetic flux is provided using blocks of Neodymium Iron Boron (NdFeB) with a typical remnant field of 1.41~T and Ni-cu-Ni coating for vacuum compatibility. Blocks are positioned on either side of the C-core gap by an Aluminium lattice. A total of 80 blocks will be used with dimensions $49\times38.5\times18$~mm, with a predicted maximum central flux density of 0.72~T. The magnet horizontal aperture is 20~mm which is the maximum size required to maintain sufficient flux density for GeV-scale use. The predicted trajectories at different electron energies, encompassing a variety of foreseen post-IP beam configurations, are shown in Fig. \ref{fig:trajectory}. 

\subsubsection{Longitudinal diagnostics}
FEBE will utilise multiple longitudinal diagnostic systems to provide measurements of relative bunch compression, RMS bunch length, and detailed longitudinal profile reconstruction. 

A bunch compression monitor based on coherent transition radiation will be installed at the end of the FEBE arc. The device uses mesh filters in order to act as a rudimentary spectrometer and provide indicative bunch duration information. The system will operate across a bandwidth of $0.1-5$~THz, measuring minimum RMS bunch lengths $\lesssim 50$~fs for charges as low as 100~pC. The detector is a bespoke pyroelectric system, employing variable gain, active noise cancellation, and Winston cones for radiation capture. Both solid and holed (for coherent diffraction radiation) targets can be used; the latter offers potential for non-invasive measurements of shot-to-shot compression jitter. Combined with accurate measurements using the upstream CLARA TDC, the bunch compression monitor will ensure maximally compressed beams are delivered to FEBE for user experiments. 

\begin{figure}[b]
\includegraphics[width=\linewidth]{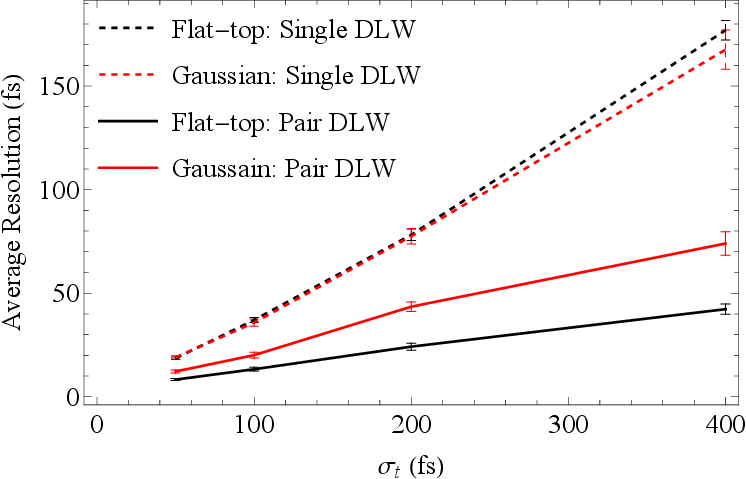}
\caption{\label{fig:dwa}Simulated average resolution of the DLW streaker at 250~pC for varying bunch length $\sigma_{t}$ for Gaussian (red lines) and flat-top (black) profiles, where flat-top has total length $4\sigma_{t}$. The improvement in resolution from using a pair of orthogonal oriented structures (solid lines) as compared to a single structure (dashed) is shown.}
\end{figure}

A passive streaker based on a dielectric wakefield (DLW) accelerator structure reported in~\cite{Saveliev2020} will be used to make bunch length measurements across the range of $\sim$50~fs to $>$ 1~ps. As compared to the previous design, the FEBE streaker uses two orthogonal waveguides mounted as a pair to compensate the effect of quadrupole-like wakefields which lead to nonlinear streaking forces; this improves the resolution of the device by a factor of $\sim$3. The simulated resolution as a function of bunch length (at 250~pC) is plotted for single and paired structure arrangements and compared in Fig.~\ref{fig:dwa}. The DLW streaker resolution does not scale favourably with reducing beam charge and will not be able to resolve the bunch length at low charges.

\subsubsection{Emittance diagnostics}
Single-shot emittance diagnostics will have high impact in novel acceleration experiments where potential beam instabilities hinder application of conventional multi-shot techniques (e.g. quadrupole scan technique). FEBE will utilise emittance diagnostics based on imaging of optical transition radiation (OTR), which is a well established technique for measuring transverse beam sizes and beam divergences. As both spatial and angular information from the electron beam is encoded in OTR, direct measurement of beam emittance is possible: this requires localising divergence measurements to discrete regions of the spatial OTR image, analogous to making an emittance measurement using a mechanical slit or ``pepper-pot.'' Two techniques are under investigation to produce an OTR based ``optical pepper-pot'': optical masking using a digital micro-mirror device (DMD)~\cite{LeSage1999}, and imaging with a micro-lens array (MLA)~\cite{Bisesto2018}.

\subsubsection{Virtual diagnostics \label{sec:virDia}}
Machine learning (ML) based algorithms can be used to predict beam parameters at a given location from a set of input machine parameters the ML model has been trained on. Such `virtual diagnostics' are particularly relevant in experiments where diagnostics cannot be installed, e.g. due to spatial and mechanical constraints, or where a non-invasive measurement is required but not available~\cite{wolfenden:23}.

As part of the FEBE design, virtual diagnostics for prediction of the IP beam size have been developed, which make an inference based on an image of the beam either up- or downstream of the IP~\cite{Wolfenden2022}. This allows, for example, non-invasive measurements of the electron beam to be performed with the FEBE 100~TW laser running through the FEC1 IP, which prevents the insertion of any screen. 

Future development will focus on virtual diagnostics for longitudinal phase space prediction, building upon recent work~\cite{Maheshwari2021} and utilising  data from the FEBE bunch compression monitor. The use of information in the form of non-invasive shot-to-shot spectral measurements has been shown to improve the accuracy of phase space predictions~\cite{Hanuka2021}, while recent work based solely on experimental data has demonstrated  predictions at significantly higher resolution to previous studies~\cite{Zhu2021}. 

\subsection{Laser system and laser transport}
The FEBE beam line will have access to a high power (100~TW: $\sim$2.5~J, $\sim$25~fs pulse duration, 5~Hz) Ti:Sapphire laser which can be combined with electron beam in FEC1. The laser requirements are compatible with the demands of plasma acceleration and can be used for ionization and/or wakefield excitation.

\begin{figure}[t]
\includegraphics[width=\linewidth]{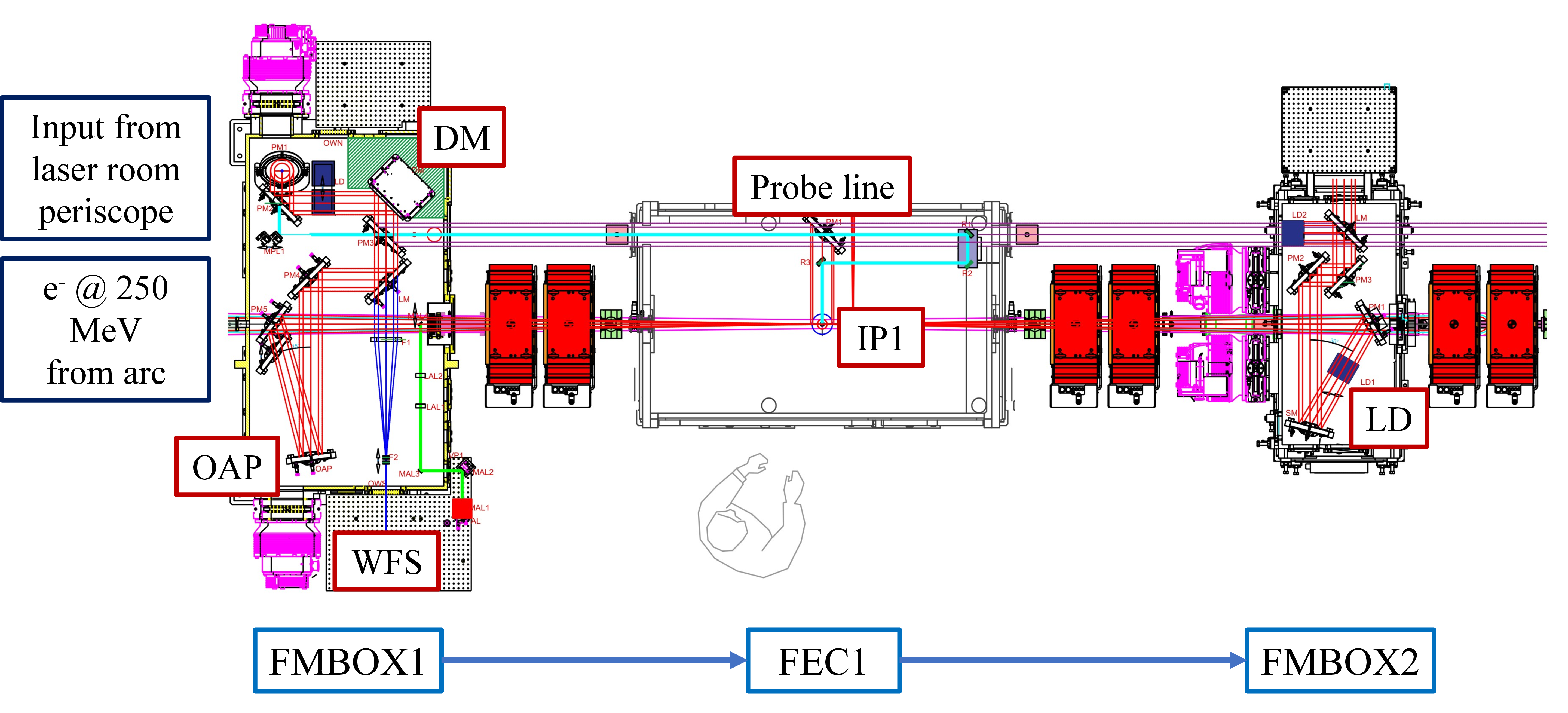}
\caption{\label{fig:laser}Laser transport within the FEBE hutch, with electrons from CLARA travelling FMBOX1-FEC-FMBOX2 (left to right). The laser is brought to a common focus with the electron beam at IP1. OAP: off-axis parabola; IP: interaction point; DM: deformable mirror; WFS: wavefront sensor; LD: laser dump.}
\end{figure}

The laser system including vacuum compressor is housed in a dedicated laser room immediately on top of the FEBE hutch. Light from the compressor is transported in vacuum from the laser room to the hutch via a radiation-shielded periscope; this allows personnel to access the laser room with electron beam in the FEBE hutch. The laser transport system inside the hutch is shown in Fig.~\ref{fig:laser} and facilitates two primary transport arrangements: 1) collinear propagation of laser light with the electron beam to a focus in FEC1, and 2) delivery of the laser without focusing directly to FEC1 for general exploitation, which can also be used for a probe line. 

Laser focusing on the co-linear path is achieved using a $\sim$3.5~m off-axis parabola housed in FMBOX1, with laser and electron beam combined on a holed mirror. This is the longest focus which can be generated while keeping the laser within the footprint of the FEBE hutch. The mirror box includes an adaptive optic system for focal spot optimisation, consisting of a deformable mirror and wavefront sensor mounted at the conjugate plane to the deformable mirror. Leakage paths into air will be utilised for other laser diagnostics, including shot-by-shot measurements of pulse energy, spectrum and beam pointing. The laser is allowed to expand after the focus to a safe intensity before being separated from the electron beam using a holed mirror in FMBOX2, which also includes space for laser exit mode diagnostics and laser termination.

\subsection{Timing and synchronization}
FEBE laser synchronization will be performed using systems developed for other systems on CLARA, aiming to deliver $<$10~fs laser-electron beam synchronization. The timing architecture is similar to that employed at larger x-ray FEL facilities~\cite{Schulz2015} and is split into three key systems: 
 \begin{enumerate}
     \item Ultrastable optical clock based on a commercial low-noise fibre laser system (1560~nm, 250~MHz Er/Yb fiber oscillator; Origami, OneFive), phase-locked to a rf master oscillator for long-term stability.
     \item The CLARA stabilized optical timing network   to deliver the optical clock to several clients on the accelerator, with active correction of the fiber length via measurement of the round-trip time. 
     \item End-station synchronization (rep rate locking) to the optical clock, including laser-laser (via optical cross-correlation) and laser-rf synchronization.  
 \end{enumerate}
Two stabilized links will be routed from the CLARA optical timing network to the FEBE beam line; one for synchronization of the 100~TW laser, and a second for a Bunch Arrival-time Monitor (BAM).   

To improve the synchronisation of the laser, a two-colour balanced optical cross-correlator (TC-BOXC) based on periodically-poled lithium niobate waveguide~\cite{christie:14th} has been developed. The TC-BOXC will be used to rep rate lock the FEBE Laser master oscillator to the CLARA optical timing network. Rep rate locking will be based on a hybrid locking configuration with that uses an RF mixer and photo-detector to perform coarse locking prior to locking the laser with the more sensitive TC-BOXC.

The waveguide based TC-BOXC has a measured resolution of 0.97~mV/fs and a theoretical resolution of 4.2~mV/fs. The theoretical resolution of this device provides an order of magnitude improvement over conventional bulk crystal cross correlators. The TC-BOXC is an all fibre device which is more robust against environmental fluctuations and requires less (valuable) optical table space compared to its free space counterpart.

The BAM will be based on PCB substrate with rod-shaped pick-ups in a similar design to~\cite{Scheible2021}. The initial performance target is 10~fs resolution for bunch charges as low as $<$5~pC. Initial tests are expected to be performed with the BAM mounted within FEC2. 

\subsection{Vacuum management}
The FEBE vacuum design accommodates a wide range of possible experiments. Of these experiments, those targeting plasma acceleration are the most demanding on vacuum management due to the associated gas injection at high pressure.

\begin{figure}[!t]
    \includegraphics[width=\linewidth]{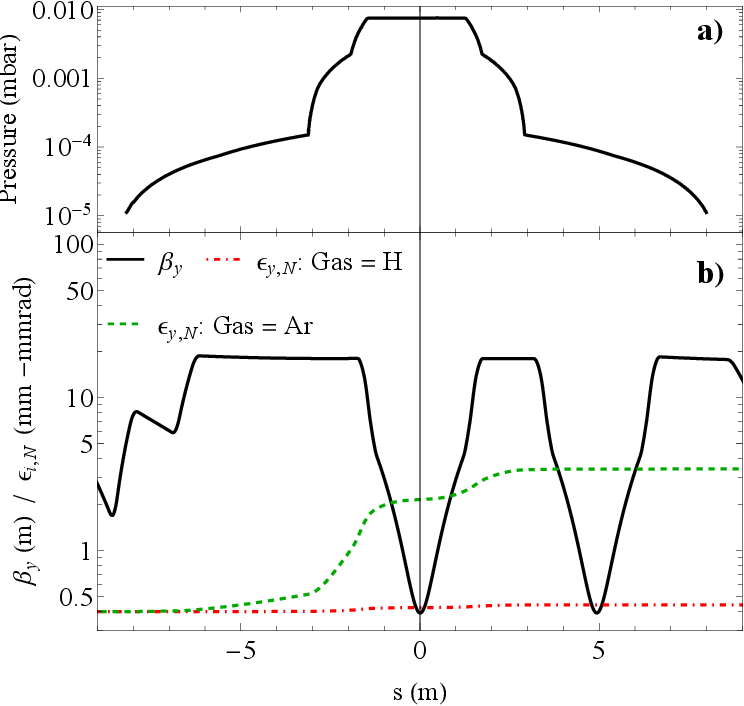}
    \caption{\label{fig:gas}(a) Gas pressure profile for Hydrogen and Argon and (b) normalised vertical emittance and vertical beta-functions $\pm$10~m from the  FEC1 IP (located at $s=0$~m). Base gas pressure of 0.01~mbar at the IP is assumed, with a pressure reduction to $10^{-6}$~mbar at either end of the vacuum line.}
\end{figure}

The FEBE design transfers vacuum management to outside of the experiment chambers by introducing aperture restrictions throughout the FEBE beam line.   This provides maximum flexibility to the user as required, however results in gas particles propagating up and downstream of the source. This increases the pressure in the FEBE arc and main CLARA accelerator above nominal levels. 

High gas pressures lead to strong beam-gas scattering effects which increase the electron beam normalised emittance. Following the work in~\cite{Pimpec2014} we have assessed the effects of high gas loading on beam emittance for representative experimental gas profiles. The approach utilised modelled Twiss parameters throughout the line via linear interpolation; the gas profile was built from a series of defined pressure points and was similarly interpolated. The incoming horizontal and vertical electron beam emittance was 5 and 0.4~$\mu$m-rad. The assessment is performed for 0.01~mbar of gas pressure at the IP (worst case estimate, based on a gas jet positioned at $z=0$~m with 100~bar backing pressure and 100~ms opening time), with various aperture restrictions in the FEC1, as well as beam apertures in FMBOX1/2 due to mirrors.

Simulations were performed using MolFlow~\cite{MolFlow}. These demonstrate $\sim10^{-6}$~mbar at the beam shutter on the outside wall of the FEBE hutch. The impact of Hydrogen and Argon gas species were compared. Fig.~\ref{fig:gas} shows the gas density profile and matched beta-functions for the FEBE beam line between the dipole magnet and the end of the hutch, and the vertical normalised emittance for both gas species. As expected, due to the higher molecular weight, there is a significant increase in emittance for Argon as compared to Hydrogen. The vertical emittance increases by almost an order of magnitude, primarily due to the combination of relatively high beta-function and gas pressure seen between FMBOX1 and the IP. Emittance blow-up inside the FEC1 chamber is minimal as the transverse beta-functions rapidly decrease in this region. Due to the significantly larger expected horizontal emittance for these experiments, $\sim$5~mm-mrad, only small changes are seen in the horizontal plane.
 
The vacuum management system will be updated following the results of initial operations and testing of gas targets. Should gas loading exceed expected values in practice, an alternative design solution based on management close to target (similar to as deployed by [\onlinecite{Delbos2018}]) will be used.

\subsection{\label{sec:ml}Longitudinal profile shaping}

Optimisation of the longitudinal profile of the photoinjector laser provides two key benefits to the electron beam optics. The first benefit is flexible variation and correction of the laser profile to improve the electron bunch longitudinal and transverse optics and mitigate space-charge and other collective effects in the beam transport. The second benefit is modification of the longitudinal profile to maximise the potential of novel acceleration experiments through creation of high transformer ratio profiles, or for the generation of drive and witness bunches in a single RF bucket. The capability to shape the longitudinal profile of electron bunches has therefore been developed for use in FEBE and, while predominantly targeted to drive-witness wakefield beam experiments, will be made available to all users. Longitudinal laser shaping is performed via control of the photoinjector laser profile,\cite{Pollard2022} and can be used either standalone or in combination with other methods described in Secs. \ref{sec:layout} and \ref{sec:s2e} above, including use of the mask installed in the FEBE arc.

\begin{figure}[b]
\includegraphics[width=8.6cm]{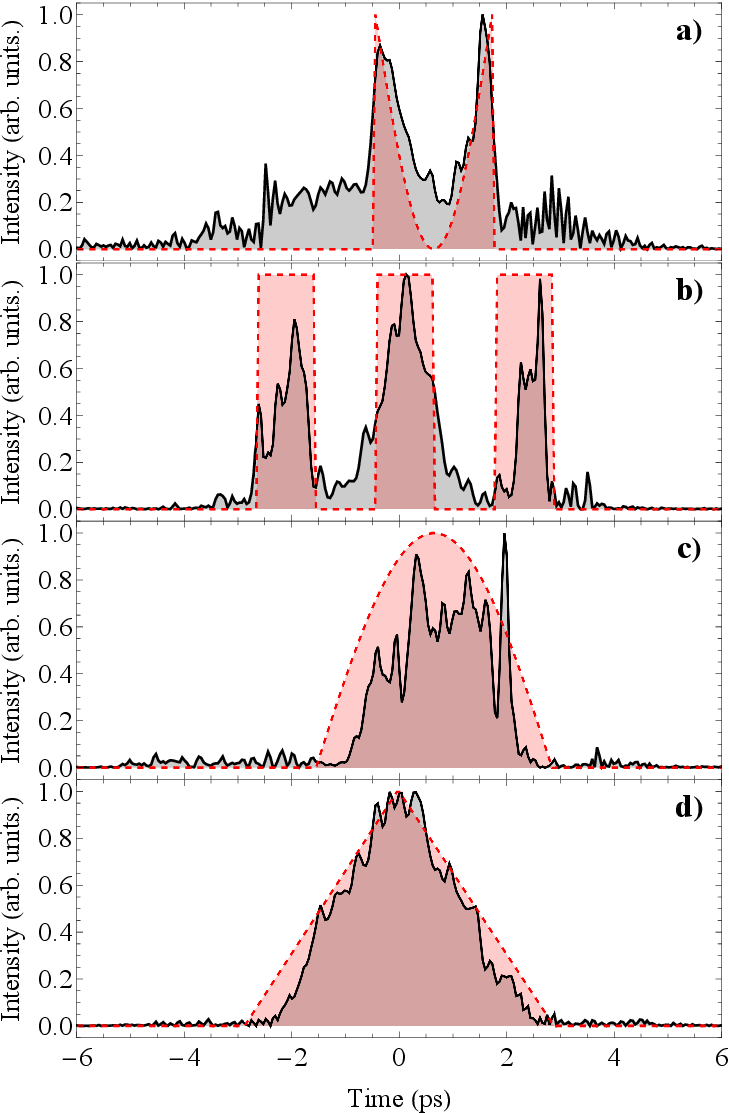}
\caption{\label{fig:ml}Demonstration of solutions to deliver arbitrary photoinjector laser temporal profiles via spectral phase manipulation, comparing target (red dashed lines) and predicted (black solid) profiles. Results for different profiles are compared, including: a) double-ramp, b) triple flat-top, c) cut sine-wave, and d) triangular.}
\end{figure}

Control of the photoinjector laser is performed using an acousto-optic modulator integrated into a 4-f spectral filter~\cite{Hillegas1994}. To shape the laser pulse temporally, the spectral phase of the pulse can be adjusted by varying the temporal phase of the acoustic wave (created using an RF pulse generator) which drives the modulator. Pulses can also be shaped temporally by varying the temporal amplitude of the acoustic wave, however this approach reduces the output pulse energy and is undesirable when maintaining high charge operation from the photoinjector. 

In order to produce a particular target pulse temporal intensity profile, a suitable spectral phase mask must first be identified.  This is nontrivial for arbitrary shapes, requiring full knowledge of both the phase and amplitude in either the spectral or temporal domain to fully define the pulse. With only temporal and the spectral intensity fully known, a suitable phase mask must be identified to fully specify the output pulse. 

For maximum flexibility and rapid customisation, a machine learning model has been developed to find the required phase mask to achieve a (user defined) target pulse temporal profile. Training data was generated from $10^{6}$ pairs of spectral phase profiles and matched temporal intensity profiles, with a further $10^{5}$ pairs generated for the test data set. To encode the physical limitation of the acousto-optic modulator bandwidth into the network, the model includes a regulariser that acts to limit the gradient of the spectral phase profile to within physical limits. High quality (and physically realisable) matches to target have been achieved for a range of pulse profiles, as shown in Fig.~\ref{fig:ml}. The system will in future be trained using live data to explore and account for expected deviations between simulation and practice.

Deployment of the ML system for control of the laser pulse shape has begun, with machine operators able to specify arbitrary pulse shapes for which the ML system produces an appropriate spectral phase profile. This generated profile is then sent to the laser control system and applied to the acousto-optic modulator, with a total time between user request and laser activation of less than 100 ms. 

\section{\label{sec:summary} Summary}
A new beam line for Full Energy Beam Exploitation (FEBE) has been designed and is currently undergoing installation on the CLARA test facility at STFC Daresbury Laboratory. The goal of this beam line is to support a wide variety of user-driven experiments utilising 250~MeV ultrabright electron bunches delivered at repetition rates up to 100~Hz. The beam line incorporates two large volume experiment chambers with a shielded user hutch, for ease of user access and flexibility in setup of novel experiment apparatus.

A key component of the foreseen future experiment programme is novel acceleration, with expressions of interest for plasma acceleration (laser and beam-driven) and structure wakefield acceleration. This has driven key components of the beam line design, including beam diagnostics for GeV-scale, 10~fs duration and micrometer-scale transverse profile electron bunches. The beam line includes the infrastructure for combining electron bunches with a high power (100~TW) laser, housed immediately above the beam line and brought into the hutch via a dedicated vacuum laser transport. The laser will be synchronized to the CLARA bunches using an optical timing architecture similar to that applies at larger x-ray FEL facilities.

CLARA will exit shutdown for installation of accelerator modules third quarter of 2023, and will proceed to enter a period of technical and machine commissioning running through to the second half of 2024; installation and commissioning of the 100~TW laser will then take place, running through to early 2025. An open call to the community is expected to be issued mid 2024 for beam time early 2025, but will be contingent on the results of machine commissioning. Future operations are expected to be divided between user access and ongoing machine development. Machine development will focus on achieving more challenging electron bunch parameters and configurations, as well as improving reliability in beam delivery. Time will also be used to ensure CLARA can continue to be a testing ground for future UK accelerator facilities, such as UK XFEL~\cite{ukxfel, dunning:2023}.

\bibliography{FEBE_APS}

\end{document}